\documentclass[graybox]{svmult}

\usepackage{mathptmx}       % selects Times Roman as basic font
\usepackage{helvet}         % selects Helvetica as sans-serif font
\usepackage{courier}        % selects Courier as typewriter font
\usepackage{type1cm}        % activate if the above 3 fonts are
\usepackage{makeidx}         % allows index generation
\usepackage{graphicx}        % standard LaTeX graphics tool
\usepackage{multicol}        % used for the two-column index
\usepackage[bottom]{footmisc}% places footnotes at page bottom

\newcommand{\msun}{\mbox{$M_{\odot}$}}
\newcommand{\lsun}{\mbox{$L_{\odot}$}}

\newcommand{\beq}{\begin{equation}}
\newcommand{\eeq}{\end{equation}}
\newcommand{\beqa}{\begin{eqnarray}}
\newcommand{\eeqa}{\end{eqnarray}}

\makeindex             % used for the subject index
\begin{document}

\title*{Very Massive Stars in the Local Universe}
\author{Jorick S. Vink}
\institute{Jorick S. Vink\at Armagh Observatory, \email{jsv@arm.ac.uk}}
\maketitle

\abstract{
Recent studies have claimed the existence of very massive stars (VMS)
up to 300\,\msun\ in the local Universe.
As this finding may represent a paradigm shift for the canonical
stellar upper-mass limit of 150\,\msun, it is timely to 
evaluate the physics specific to VMS, which is currently missing. 
For this reason, we decided to construct a book entailing 
both a discussion of the accuracy of VMS masses (Martins), as well as the physics 
of VMS formation (Krumholz), mass loss (Vink), instabilities (Owocki), evolution (Hirschi), 
and fate (theory -- Woosley \& Heger; observations -- Smith).}

\section{Introduction}
\label{sec:intro}

It has been thought for many years that very massive stars (VMS) with masses
substantially larger than 100\,\msun\ may occur more frequently in the
early Universe, some few hundred million years after the Big Bang. 
The reason for the expectation that the first few stellar generations 
would generally have been more massive is that there 
was less cooling during the formation process of these metal-poor objects 
than in today's metal-rich Universe 
(e.g. Bromm et al. 1999; Abel et al. 2002; Omukai \& Palla 2003; Yoshida et al. 2004; Ohkubo et al. 2009). 

Furthermore, as radiation-driven winds are thought to be weaker at the lower metal content 
of the early Universe (e.g. Kudritzki 2002; Vink \& de Koter 2005; Krticka \& Kubat 2006; Gr\"afener \& Hamann 2008; Muijres et al. 2012), 
this could imply 
that the final masses of VMS in the early Universe 
would be almost equally high as their initial masses. This could then lead 
to the formation of $10^2 - 10^3\, \msun$ 
intermediate-mass black holes (IMBHs), with masses 
in between stellar mass black holes and supermassive black
holes of order $10^5\,\msun$ in the centres of galaxies. IMBHs have been hypothesized  
to be the central engines of the ultraluminous x-ray sources (ULXs).  
Moreover, in a high stellar mass -- low mass loss -- situation it might become possible  
to produce pair-instability supernovae (PISNe) in the initial mass range of 140-260\,\msun\ 
(Woosley et al. 2002; see also Fowler \& Hoyle 1964; Barkat et al. 1967; Bond et al. 1984; 
Langer et al. 2007; Moriya et al. 2010; Pan et al. 2012; Dessart et al. 2013; Whalen et al. 2013).  
Such PISNs are very special as just one such explosion could potentially produce more 
metals than an entire initial mass function (IMF) below it (Langer 2012).

Interestingly, Crowther et al. (2010) re-analyzed
the most massive hydrogen-and nitrogen-rich Wolf-Rayet (WNh)
stars in the center of R136, the
ionizing cluster of the Tarantula nebula in the Large Magellanic
Cloud (LMC). The conclusion from their analysis was that stars usually
assumed to be below the canonical stellar upper-mass limit of
150\,\msun\ (of e.g. Figer 2005),
were actually found to be much more luminous (see also Hamann et al. 2006; Bestenlehner et al. 2011),
with initial masses up to $\sim$200-300\,\msun.
As this finding may represent a paradigm shift for the canonical
stellar upper-mass limit of 150\,\msun, it is timely to discuss
the status of the data as well as VMS theory. 

Whilst textbooks and reviews have been devoted to the physics of canonical 
massive single and binary stars (Maeder 2009; Langer 2012)
there is as yet no source that specifically 
addresses the physics unique to VMS. 
As such objects are in close proximity to the Eddington limit, this is 
likely to affect both their formation, and via their mass loss also their fates. 

\section{The role of very massive stars in the Universe}

The first couple of stellar generations may be good candidates for the reionization of the 
Universe (e.g. Haehnelt et al. 2001; Barkana \& Loeb 2001; Ciardi \& Ferrera 2005; Fan et al. 2006) and 
their ionizing properties  at very low metallicity ($Z$) may also be able to explain the 
extreme Ly$\alpha$ and He {\sc ii} emitting galaxies at high redshift 
(Malhotra \& Rhoads 2002; Kudritzki 2002; Schaerer 2003; Stark et al. 2007; Ouchi et al. 2008).

Notwithstanding the role of the first stars, the interest in
the current generation of massive stars has grown as well.
Massive stars are important drivers for the evolution
of galaxies, as the prime contributors to the chemical and
energy input into the interstellar medium (ISM) through
stellar winds and supernovae (SNe).
A number of exciting developments have taken place in recent years, including the 
detection of long-duration gamma-ray bursts (GRBs) at redshifts of 9 (e.g. Tanvir et al. 2009), just
a few hundred millions years after the Big Bang
(Cucchiara et al. 2011). This provides convincing
evidence that massive stars are able to form and die massive when
the Universe was not yet enriched.

Very massive stars are usually found in and around young massive clusters, such as the
Arches cluster in the Galactic centre and the local starburst
region R136 in the LMC. 
Young clusters are also relevant for
the unsolved problem of massive star formation.
For decades it was a real challenge to form stars over $10-20\,\msun$, as radiation
pressure on dust grains might halt and reverse the accretion flow onto
the central object (e.g. 
Yorke \& Kruegel 1977; Wolfire \& Cassinelli 1987). Because of this
issue, theorists have been creative in forming massive stars
via competitive accretion and collisions in dense cluster environments
(e.g., Bonnell et al. 1998).  In more recent times several multi-D
simulations have shown that massive stars might form via disk
accretion after all (e.g., Krumholz et al. 2009; Kuiper et al.  2010).
In the light of recent claims for the existence of VMS
in dense clusters, however, the issue of
forming VMS in extreme environments is discussed by Mark Krumholz in Chapter 3.

The fact that so many VMS are located within dense stellar clusters still allows for an 
intriguing scenario in which VMS may originate from 
collisions of smaller objects (e.g., Portegies Zwart et al.
1999; G\"urkan et al. 2004),
leading to the formation of VMS up to 1000\,\msun\ at the
cluster center, which may produce IMBHs at the end of their lives, but only
if VMS mass loss is
not too severe (see Belkus et al. 2007, Yungelson et al. 2008, Glebbeek et al. 2009, Pauldrach
et al. 2012).

\section{Definition of a very massive star}
\label{sec:def}

One of the very first questions that arises when one prepares a book on VMS is 
what actually constitutes
a ``very'' massive star. One may approach this in several different ways.

Theoretically, ``normal'' massive stars with masses above $\sim$8\,\msun\ are those that
produce core-collapse SNe (Smartt et al. 2009), but what happens at the upper-mass end?
Above a certain critical mass, one would expect the occurrence of PISNe, and ideally
this could be the lower-mass limit for the definition of our VMS.
However, in practice this number is not known a priori (due to mass loss), and therefore 
the initial and final masses are likely not the same. In other words, the initial main-sequence
mass for PISN formation is model-dependent, and thus somewhat arbitrary.
Furthermore, there is the complicating issue of pulsational pair-instability (PPI) 
at masses below those of full-fedged PISNe
(e.g. Woosley et al. 2007). One could alternatively 
resort to the mass of the helium (He) core for which stars
reach the conditions of electron/positron pair-formation instability.
Heger showed this minimum mass to be $\sim 40\,\msun$ to encounter the 
PPI regime and $\sim 65\,\msun$ to enter the arena of the true 
PISNe (see also Chatzopoulus \& Wheeler 2012).

Another definition could involve the spectroscopic
transition between normal main-sequence O-type
stars and hydrogen-rich Wolf-Rayet stars (of WNh type), which have also been shown
to be core H burning main sequence objects.
However, such a definition would be dependent on the mass-loss 
transition point between O-type and WNh stars, which is set by
the transition luminosity (Vink \& Gr\"afener 2012) and is expected 
to be $Z$ dependent.

For these very reasons, we decided at the joint discussion meeting 
at the 2012 IAU GA in Beijing to follow a more pragmatic approach, 
defining stars to be {\it very} massive when their initial masses are $\simeq 100\,\msun$ (Vink et al. 2013).

\section{The very existence of very massive stars}
\label{sec:exist}

With this definition, the question of whether {\it very} massive stars exist can
easily be answered affirmatively, but the more relevant question during the joint discussion 
was whether the widely
held ``canonical'' upper-mass limit of $150\,\msun$ has been superseded, as 
some part of the astronomical community had expressed some skeptism regarding very 
high masses in R136, in the light of an earlier 
spectacular claim for the existence of a $2500\,\msun$ star R136 in the 30 Doradus region
of the LMC (e.g. Cassinelli et al. 1981).
Higher spatial resolution showed that R136 was actually not a single supermassive star, but
it eventually revealed a young cluster containing several lower mass objects, including
the current record holder R136a1.

Over the last few decades there has been a consensus of a 150\,\msun\ stellar upper mass limit
(Weidner \& Kroupa 2004; Figer 2005; Oey \& Clarke 2005, Koen 2006),
albeit the accuracy of these claims was surprisingly low (e.g. Massey 2011).
Crowther et al. (2010) re-analyzed the VMS data
in R136 claiming that the cluster hosts several stars with masses as high
as 200-300\,\msun. In addition they performed a sanity check on similar WNh objects
in the Galactic starburst cluster NGC\,3603.
Although these objects were fainter than
those in R136, the advantage was the available dynamical mass estimate by
Schnurr et al. (2008) of the binary object NGC\,3603-A1 with a primary mass of 
$116 \pm 31 \msun$.
This was deemed important as the least model-dependent way to obtain stellar masses is
through the analysis of the light-curves and radial velocities
induced by binary motions (see Martins' Chapter 2). 

It could still be argued that
the luminosities derived by Crowther et al. are uncertain
and that these central WNh stars might in reality involve multiple sources due 
to insufficient spatial resolution, especially considering that  
the highest resolution data of the young Galactic 
Arches cluster with the largest telescope (Keck) only has a limiting 
resolution of 50 milli-arcsec, and 
given that R136 is 7 times more distant than the Arches cluster, the 
achievable resolution if the Arches cluster were in the LMC would mean 
that R136 would not be resolved.
This suggests that we still cannot be $100\%$ certain that the bright 
WNh stars in R136 could not ``break up'' into lower-mass objects.

For this reason it was rather relevant that Bestenlehner et al. (2011) found 
an almost identical twin of R136a3 WNh star in 30 Doradus: VFTS\,682.
Its key relevance is that it is located 
in apparent isolation from the R136 cluster, and as a result
the chance of line-of-sight contamination is insignificant in comparison to R136.
The VFTS\,682 object thus offered a second sanity check on the reliability of the
luminosities of the R136 core stars.
Bestenlehner et al. argued for a high luminosity of log($L/\lsun$) $= 6.5$ with a
present-day mass of 150\,\msun\ for VFTS\,682, which implies
an initial mass on the zero-age main sequence (ZAMS) 
higher than the canonical upper-mass limit.

In other words, although one cannot exclude the possibility that the 
object R136a1 claimed to be $\sim$300\,\msun\ in
the R136 cluster might still ``dissolve'' when higher spatial resolution observations 
become available, the sanity checks involving binary dynamics and isolated objects make it
quite convincing that stars with ZAMS masses at least up to $200\,\msun$ exist.

A more detailed overview of the masses of VMS and the upper end of the IMF will be described in Martins' Chapter 2.

\section{The evolution and fate of very massive stars}

Very massive stars are thought to evolve almost chemically
homogeneously (Hirschi's Chapter 6), implying that knowing
the exact details of the mixing processes (e.g.,
rotation, magnetic fields) are less relevant in comparison
to their canonical $\sim$10-60\,\msun\
counterparts. Instead, the evolution and death of VMS is
dominated by mass loss.

At some level it does not matter 'how' VMS 
became such massive objects. First of all we do not yet definitively know the 
formation mode of 'very' massive stars, and whether the formation involves
disk accretion or coalescence of less massive objects. 
Secondly, there is a possibility that binary evolution already 
during early core hydrogen (H) burning resulted in the formation of massive 
blue stragglers (Schneider et al. 2014; de Mink et al. 2014), but the fate of 
these effectively single VMS will naturally be determined by single-star mass loss. 

The existence of the Humphreys-Davidson (HD) limit at approximately solar metallicity 
tells us that VMS do not become red supergiants (RSG) but that they
remain on the hot side of the Hertzsprung-Russell (HR) diagram as 
luminous O stars and Wolf-Rayet-type 
objects. For these hot stars the mass loss is thought to be driven by 
million of iron lines in a radiatively-driven wind,
but what is not yet known is whether episodes of super-Eddington (Shaviv 1998), 
continuum-driven mass loss (such as may occur
in Eta Carinae and other
Luminous Blue Variable (LBV) star eruptions) may also play 
a role (see Vink's Chapter 4 and Owocki's Chapter 5). 
What is clear is that the Eddington $\Gamma$ limit will play 
a dominant role in the mass-loss physics.

We should also note that the Eddington limit is relevant for another 
issue relating to VMS physics. 
When objects approach the Eddington limit, they may or may not {\it inflate} 
(Ishii et al. 1999; Petrovic et al. 2006), i.e. 
be subject to enormous radius and temperature changes (Gr\"afener et al. 2012). 
This implies that the temperatures and thus the ages of VMS are highly uncertain.

A final issue concerns the fate of VMS. In the traditional view, after core H-burning, VMS would 
become LBVs, remove large amounts of mass, exposing their bare-naked helium (He) cores, burn
He for another $10^{5}$ before giving rise to H-poor Type Ibc SNe (e.g. Conti 1976; Yoon et al. 2012; Georgy et al. 2012).
However since 2006 there have been indications that some massive stars 
may explode prematurely as H-rich type II SNe already during the LBV 
phase (Kotak \& Vink 2006; Gal-Yam et al. 2007; Mauerhan et al. 2013).

Might some of the most massive stars even produce PISNe? And how do PISNe compare to the general
population of super-luminous SNe (SLSNe) that have recently been
unveiled by Quimby et al. (2011), and are now seen out to high redshifts
(Cooke et al. 2012)? 
Gal-Yam et al. (2009) discovered an intruiging 
optical transient with an observed light curve that
fits the theoretical one calculated from pair-instability supernova with a He core mass 
around $100\,\msun$ (see also Kozyreva et al. 2014). 

Even if the SLSNe turn out to be unrelated to PISNe as argued by Nicholl et al (2013) and Inserra et al. (2013),
we should note that alternative models such as magnetar models (e.g. Kasen \& Bildsten 2010)
would also involve rather massive stars, and if the high luminosity is 
not the result of a magnetar, but 
for instance due to mass loss, then the amounts of mass loss inferred for interacting 
type IIn SNe are so humongous (of order tens of solar masses; see Smith's Chapter 8) that 
they can only originate from VMS.

In summary, the evolution of VMS into the PISN and/or SLSNe regime 
can only be understood once we obtain a comprehensive framework regarding 
the evolution and physics of VMS. In this book, a number of 
experts discuss aspects of their research 
field relevant to VMS in the local Universe.  
In Chapter 2 Fabrice Martins discusses the observational data of VMS with a
special emphasis on the luminosity and mass determinations of both single and 
binary VMS. 
The rest of the book is mostly theoretical. In Chapter 3, Mark Krumholz discusses 
the different formation modes of VMS. As mass loss is  
so dominant for the evolution and fate of VMS, the next topics involve 
the physics of both their stellar winds (Jorick Vink; Chapter4) and instabilities (Stan 
Owocki; Chapter 5), before
Raphael Hirschi discusses the evolution of VMS in Chapter 6. We finish with an 
overview of the possible theoretical outcomes in Chapter 7 by Woosley \& Heger, and 
an overview of the observations of VMS fate by Nathan Smith in Chapter 8.

{}

\end{document}